\documentclass[prd,aps,
twocolumn,superscriptaddress,preprintnumbers,nofootinbib]{revtex4-1}
\usepackage{pslatex}
\usepackage[pdftex]{graphicx}
\usepackage{psfrag}
\usepackage{epsfig}
\usepackage{color}
\usepackage{cancel}
\usepackage{slashed}
\usepackage{amssymb}
\usepackage{amsmath}
\usepackage{hyperref}

\begin{document}
\preprint{ULB-TH/14-25}

\title{Dynamical flavor origin of $\mathbb{Z}_N$ symmetries}

\author{D. Aristizabal Sierra}%
\email{daristizabal@ulg.ac.be}%
\affiliation{IFPA, Dep. AGO, Universit\'e de Li\`ege, Bat B5, Sart
  Tilman B-4000 Li\`ege 1, Belgium}
\author{Mika\"el Dhen}%
\email{mikadhen@ulb.ac.be}%
\affiliation{Service de Physique Th\'eorique, Universit\'e Libre de
  Bruxelles, Boulevard du Triomphe, CP225, 1050 Brussels, Belgium}
\author{Chee Sheng Fong}%
\email{fong@if.usp.br}%
\affiliation{Instituto de F\'isica, Universidade de S\~ao Paulo,
  C. P. 66.318, 05315-970 S\~ao Paulo, Brazil}
\author{Avelino Vicente}%
\email{Avelino.Vicente@ulg.ac.be}%
\affiliation{IFPA, Dep. AGO, Universit\'e de Li\`ege, Bat B5, Sart
  Tilman B-4000 Li\`ege 1, Belgium}

\begin{abstract}
  Discrete Abelian symmetries ($\mathbb{Z}_N$) are a common
  ``artifact'' of beyond the standard model physics models.  They
  provide different avenues for constructing consistent scenarios for
  lepton and quark mixing patterns, radiative neutrino mass generation
  as well as dark matter stabilization. We argue that these symmetries
  can arise from the spontaneous breaking of the Abelian $U(1)$
  factors contained in the global flavor symmetry transformations of
  the gauge invariant kinetic Lagrangian. This will be the case
  provided the ultra-violet completion responsible for the Yukawa
  structure involves scalar fields carrying non-trivial $U(1)$
  charges. Guided by minimality criteria, we demonstrate the viability
  of this approach with two examples: first, we derive the
  ``scotogenic'' model Lagrangian, and second, we construct a setup
  where the spontaneous symmetry breaking pattern leads to a
  $\mathbb{Z}_3$ symmetry which enables dark matter stability as well
  as neutrino mass generation at the 2-loop order. This generic
  approach can be used to derive many other models, with residual
  $\mathbb{Z}_N$ or $\mathbb{Z}_{N_1}\times \cdots \times
  \mathbb{Z}_{N_k}$ symmetries, establishing an intriguing link
  between flavor symmetries, neutrino masses and dark matter.
\end{abstract}

\maketitle
\section{Introduction}
\label{sec:intro}%
Addressing the flavor puzzle, that is to say pinning down the origin
of the standard model (SM) fermion mass hierarchies and mixing
patterns, has led to the formulation of different and ample number of
theoretical ideas. Although seemingly unrelated, most of these
approaches follow two conceptually distinct theoretical trends: $(a)$
the underlying flavor theory involves new flavor symmetries under
which different generation SM fermions (quarks and charged leptons)
carry different charges. Mass hierarchies and mixing patterns are thus
understood as a consequence of the different transformation properties
of quarks and charged leptons, which in the SM Yukawa sector---being
an effective realization of the fundamental flavor theory---are not
manifest. Certainly, the Froggatt-Nielsen mechanism
\cite{Froggatt:1978nt} provides the most representative example for
this kind of approaches.  $(b)$ The other avenue consists in promoting
the maximal global flavor symmetry of the SM gauge invariant kinetic
Lagrangian ($G_F=U(3)^5\times U(1)_H$, in shorthand notation)
\cite{Chivukula:1987py} to a fundamental flavor symmetry, something
that calls for a $G_F$-invariant ultra-violet (UV) completion, endowed
with scalar (flavon) fields capable of triggering spontaneous symmetry
breaking (SSB) of $G_F$ (or its subgroups)
\cite{Wilczek:1978xi,Anselm:1996jm,Berezhiani:2001mh,Feldmann:2009dc,Albrecht:2010xh,Grinstein:2010ve,Alonso:2011yg,Nardi:2011st,Mohapatra:2012km,Espinosa:2012uu,Fong:2013dnk,Alonso:2012fy,Alonso:2013mca,Alonso:2013nca}. In
this picture, therefore, mass (Yukawa) hierarchies and mixing patterns
result from a high-scale dynamics rather than by a mismatch in SM
fermion fields new quantum numbers.

Arguably, if a theory of flavor is indeed at work at a certain
high-energy scale ($\Lambda$), it should provide as well the framework
for other phenomenological puzzles to be addressed. In particular, one
could envisage a more profound picture from where neutrino masses, the
baryon asymmetry of the universe and dark matter emerge. Thus, from
this perspective, in principle unrelated phenomena would be just
manifestations of a comprehensive scenario in which the flavor
symmetry plays---directly or indirectly---an essential r\^ole in e.g.
neutrino mass generation or dark matter stabilization.

In this letter we show that several aspects of such endeavor can be
pursued without specifying the complete UV theory. Adopting approach
$(b)$, we consider the generation of residual discrete Abelian
symmetries $\mathbb{Z}_N$ whose origin can be traced back to the
flavor symmetry of the gauge invariant kinetic Lagrangian. We rely on
the Abelian part of the complete flavor symmetry, following
well-established methods to induce residual symmetries from
$U(1)$-invariant theories \cite{Krauss:1988zc,Petersen:2009ip}. The
feasibility of the approach is shown with two example models, both
with an extended lepton sector and a second Higgs doublet: (i) a
minimal scenario where the breaking of the flavor symmetry leads to
the {\it scotogenic} model Lagrangian \cite{Ma:2006km}, and (ii) a
different scenario leading to a remnant $\mathbb{Z}_3$ symmetry, which
not only stabilizes dark matter, but also guarantees neutrino mass
generation at the 2-loop level. These worked-examples are of course
not unique, e.g.  several other higher order remnant symmetries can be
generated, including even direct products $\mathbb{Z}_{N_1}\times
\cdots \times \mathbb{Z}_{N_k}$.

SSB of $U(1)$ symmetries and their connection with discrete
$\mathbb{Z}_N$ symmetries are not at all new subjects
\cite{Walker:2009en,Batell:2010bp,Lindner:2013awa}. We find however a
pivotal conceptual difference between these approaches and what we
here aim at discussing: the $U(1)$ factors are ``sourced'' by the same
symmetry that dictates fermion mass hierarchies and mixing
patterns. Thus, in scenarios where the remnant discrete $\mathbb{Z}_N$
symmetry plays a r\^ole in e.g. dark matter stabilization and/or
neutrino mass generation, this approach can
be---conceivably---understood as a first step towards the
establishment of a common comprehensive framework for flavor, neutrino
and dark matter physics\footnote{Dark matter stabilization in the
  minimal flavor violating \cite{D'Ambrosio:2002ex} context has been
  considered in Ref.~\cite{Batell:2011tc}. Rather than making use of
  the Abelian symmetries in the flavor symmetry group, this approach
  relies on the non-Abelian part. Dark matter stabilization through
  flavor discrete symmetries was investigated in
  \cite{Hirsch:2010ru}.}.

\section{The standard model flavor symmetry and 
its remnant $\mathbb{Z}_N$ symmetries}
\label{sec:setup}%
As has been previously anticipated, the group of global symmetry
transformations of the gauge invariant kinetic terms of the SM quark
doublet ($q_L$) and singlets ($u_R, d_R$), lepton doublet ($\ell_L$)
and singlet ($e_R$), and Higgs doublet ($H$) is given by
\cite{Chivukula:1987py}
\begin{equation}
  \label{SM-global-symmetry-no-Yuk}
  G_F=\left[\prod_a SU(3)_{a}\times U(1)_{a}\right]\times U(1)_{H} 
\end{equation}
where $a=\left\{q_L,u_R,d_R,\ell_L,e_R\right\}$. The SM Yukawa
interactions, however, explicitly break this symmetry leaving behind
just five global $U(1)$ factors:
\begin{equation}
  \label{SM-global-symmetry}
  G_{SM}=U(1)_B
  \times\left[\prod_{\alpha=\{e,\mu,\tau\}}U(1)_{L_\alpha}\right]
  \times U(1)_Y\ ,
\end{equation}
readily identifiable with conservation of baryon ($B$) and lepton
flavor ($L_\alpha$) numbers (SM accidental symmetries) and hypercharge
($Y$), which according to Tab. \ref{tab:table-1} are given by the
following linear combinations of the $U(1)_a$ charges:
\begin{eqnarray}
  \label{SM-remnant-U(1)s}
  & & B = \frac{1}{3} \left( Q_{q_L} + Q_{u_R} + Q_{d_R} \right) \,, \;\;\;
  L_\alpha = \left(Q_{\ell_{L}} + Q_{e_{R}} \right)_\alpha \,, \\
  & & \!\!\!\! Y = \frac{1}{6}\left(Q_{q_L} + 4Q_{u_R} 
  - 2Q_{d_R} \right) + \frac{1}{2}\left(Q_H - Q_{\ell_L} - 2 Q_{e_R}\right)\,.
\end{eqnarray}

It is worth stressing at this point that, since massive neutrinos is
an experimental fact, $U(1)_{L_\alpha}$ is actually broken.  If
neutrinos have Dirac masses, one can have total lepton number $U(1)_L$
conservation. On the other hand, if neutrinos have Majorana masses,
even $U(1)_L$ is broken. Formal invariance of the full Lagrangian
under $G_F$ can however be recovered, provided the Yukawa couplings
are promoted to complex scalar fields, i.e. flavon fields with
definitive $G_F$ transformations (see Tab. \ref{tab:table-1}) but
singlets under the SM gauge symmetry. At the ``fundamental'' level,
this means that at some unknown---but certainly large---energy scale,
$G_F$ is an exact symmetry of the UV Lagrangian. The flavor symmetry
is then spontaneously broken by the vacuum expectation values (vevs)
of new heavy scalar degrees of freedom with suitable $G_F$
transformation properties. Thus, in that picture the SM Yukawa
Lagrangian (which emerges once below the characteristic UV energy
scale $\Lambda$ the heavy degrees of freedom are integrated out) is an
effective manifestation of the flavored UV theory, namely
\begin{equation}
  \label{MFV-lagrangian}
  -{\cal L}_\text{SM}=
  \overline{ q_{L}} \, \frac{\langle Y_u\rangle}{\Lambda}\, u_{R} \, \tilde{H}
  + \overline{q_{L}} \, \frac{\langle Y_d\rangle}{\Lambda}\, d_{R}\, H 
  + \overline{\ell_{L}} \, \frac{\langle Y_e\rangle}{\Lambda}\, e_{R} \, H 
  + \text{H.c.}\ ,
\end{equation}
where $\tilde H = \epsilon\, H^*$ with $\epsilon = i\tau_2$ and
$\tau_2$ the second Pauli matrix. The Lagrangian, written in this way,
assumes that $\langle Y_X\rangle$ triggers not only SSB of the
non-Abelian sector of $G_F$ but also of the Abelian sector, the $U(1)$
factors.  This choice is to some extent arbitrary, and indeed more
interesting possibilities do exist, see for example
\cite{Nardi:2011st}.
\begin{table}
  \centering
  \begin{tabular}{|l|c|c|c|c|c|}
    \hline
    &$\prod_a SU(3)_a$& $Q_{q_L}$ &$Q_{u_R}$  &$Q_{d_R}$
    \\\hline \hline
    $q_L$	   & $3\times1\times1\times1\times1$     & 1 & 0 & 0
    \\ \hline
    $u_R$	   & $1\times3\times1\times1\times1$     & 0 & 1 & 0
    \\ \hline 
    $d_R$	   & $1\times1\times3\times1\times1$     & 0 & 0 & 1
    \\ \hline 
    $\ell_L$ & $1\times1\times1\times3\times1$     & 0 & 0 & 0
    \\ \hline
    $e_R$	   & $1\times1\times1\times1\times3$     & 0 & 0 & 0
    \\ \hline\hline
    $H$	   & $1\times1\times1\times1\times1$     & 0 & 0 & 0
    \\ \hline 
    $Y_u$	   & $3\times\bar3\times1\times1\times1$ & 1 & $-1$ & 0
    \\ \hline
    $Y_d$	   & $3\times1\times\bar3\times1\times1$ & 1 & 0 & $-1$
    \\ \hline
    $Y_e$	   & $1\times1\times1\times3\times\bar3$ & 0 & 0 & 0
    \\ \hline
  \end{tabular}
  \begin{tabular}{|c|c|c|}
    \hline
    $Q_{\ell_L}$  & $Q_{e_R}$ & $Q_{H}$
    \\  \hline \hline
    0           & 0 & 0
    \\ \hline
    0           & 0 & 0
    \\ \hline 
    0           & 0 & 0
    \\ \hline 
    1           & 0 & 0
    \\ \hline
    0           & $1$ & 0
    \\ \hline\hline
    0           & 0 & 1
    \\ \hline
    0           & 0 & 1
    \\ \hline
    0           & 0 & $-1$
    \\ \hline
    1           & $-1$ & $-1$
    \\ \hline
  \end{tabular}
  \caption{Flavor transformation properties of the SM fermion and
    Higgs fields as well as of the scalar fields $Y_{u,d,e}$. $Q_a$
    and $Q_H$ stand for the charges of the different $U(1)$ factors.}
  \label{tab:table-1}
\end{table}

The picture described so far is however incomplete since neutrino
masses must be generated. Here we will assume that neutrinos are
Majorana fermions, thus breaking lepton number. Although one could
well extend the SM Lagrangian to include the dimension five lepton
number breaking operator $\ell\ell HH$ \cite{Weinberg:1979sa}, we
stick to the {\it standard} picture involving \emph{three}
right-handed neutrinos $\nu_R$. This implies enlarging $G_F$ to
$G_F\times U(1)_{\nu_R}\times SU(3)_{\nu_R}$
\cite{Cirigliano:2005ck,Alonso:2013mca}, and introducing additional
flavon fields to give rise to the observed mass spectrum and mixing in
the lepton sector through SSB. Furthermore, it requires extending the
definition of lepton number in Eq. \eqref{SM-remnant-U(1)s} to
$L_\alpha = \left(Q_{\ell_{L}} + Q_{e_{R}} +
  Q_{\nu_{R}}\right)_\alpha$. In this context, $U(1)_{\nu_R}$ breaking
has been shown to have deep implications, allowing the detachment of
the right-handed neutrino mass and lepton number-breaking scales
\cite{Alonso:2011jd}, or rendering right-handed neutrino production
viable, and in some cases implying large charged lepton
flavor-violating effects \cite{Sierra:2012yy}.

Certainly the dynamics of the non-Abelian sector of $G_F$ plays a
r\^ole. Actually, in full generality, one should expect the UV
completion to involve not only SSB of the non-Abelian structure via
``Yukawa'' fields, but also of the Abelian one through flavon fields.
Thus, the question is then whether that dynamics leaves traces beyond
those that we have discussed. This might be indeed the case, provided
the flavored UV completion involves scalar fields with suitable
charges under some of the $U(1)$ global factors (a single one
suffices). Let us discuss this in more detail.  Consider a simple
model of two self-interacting scalar fields ($\sigma_{1,2}$) subject
to the following global $U(1)$ transformations
\cite{Krauss:1988zc,Petersen:2009ip}:
\begin{equation}
  \label{eq:u-1-transform}
  \sigma_1 \to e^{i N \alpha}\sigma_1
  \quad\text{and}\quad
  \sigma_2 \to e^{-i\alpha}\sigma_2\ .
\end{equation}
The $U(1)$-invariant renormalizable as well as non-renormalizable
Lagrangian describing such system is given by:
\begin{equation}
  \label{eq:lag-scalar-theory}
  {\cal L} = \mu_{i}^2\sigma_i^*\sigma_i
  + \lambda_{ij}(\sigma_i^*\sigma_i)(\sigma_j^*\sigma_j)
  + \frac{\lambda_{\sigma_M}}{\Lambda^{M(N+1)-4}}\sigma_1^M(\sigma_2^N)^M\ ,
\end{equation}
with $\mu_i$ dimension one and $\lambda_{ij}, \lambda_{\sigma_M}$
dimensionless couplings.  If $\sigma_1$ acquires a vev the resulting
Lagrangian will involve a collection of terms $(\sigma_2^N)^M$, thus
being $\mathbb{Z}_N$ invariant, namely
\begin{equation}
  \label{eq:zn-invariance}
  \sigma_2\to \eta^n_N \, \sigma_2\quad\text{with}\quad
  \eta_N = e^{2\pi i/N}\quad(n=0,1,\dots, N-1)\ .
\end{equation}

Although discussed in a rather simple context, this idea can be
extended to realistic models involving fermion fields and therefore
Yukawa terms. There is however something that one should bear in
mind. SSB of the global flavor symmetry, including its $U(1)$ factors,
imply the presence of massless Nambu-Goldstone bosons for which a
large variety of phenomenological constraints exist, including rare
decays, cosmological and astrophysical data. One solution is that of
gauging part of or the full flavor symmetry, which of course implies
the presence of new gauge bosons and calls for gauge anomaly
cancellation which requires the introduction of new fermions
\cite{Albrecht:2010xh,Grinstein:2010ve}.  If one insists on a global
symmetry, phenomenological consistency can be achieved provided SSB
takes place at a rather high energy scale, which will suppress
couplings to fermions \cite{Wilczek:1982rv,Chang:1984ip}. However,
scalar couplings with the Higgs sector at the renormalizable level can
potentially affect the Higgs phenomenology (e.g. giving rise to Higgs
invisible decays).

\section{Emerging $\mathbb{Z}_N$-based  
  models: benchmark examples}
\label{sec:emerging-disc-symm}
Guided by the aforementioned considerations, we now turn to the
discussion of some specific $\mathbb{Z}_N$-based realizations.  Since
we consider only the (extended) lepton sector, we assume that all
beyond the SM (BSM) fields transform trivially under the quark flavor
symmetry, $U(3)_{q_L}\times U(3)_{u_R}\times U(3)_{d_R}$, and we only
specify their transformation properties under
$(SU(3)_{\ell_L},SU(3)_{e_R},SU(3)_{\nu_R})_{(U(1)_{\ell_L},U(1)_{e_R},U(1)_{\nu_R},U(1)_H)}$. In
particular, we focus on fields which give rise to light neutrino
masses.  For instance, the SM lepton and Higgs doublet are written
respectively as $\ell_L = (3,1,1)_{(1,0,0,0)}$ and
$H=(1,1,1)_{(0,0,0,1)}$.  For definiteness, we consider models
extended with a second Higgs doublet $\Phi=(1,1,1)_{(0,0,-1,-1)}$ and
three right-handed neutrinos $\nu_R=(1,1,3)_{(0,0,1,0)}$ while the
flavon fields which we will introduce below are singlets under the SM
gauge symmetry.  It is worth stressing that the approach discussed
here is not limited to scenarios with right-handed neutrinos. In
principle, it can be implemented in a large variety of BSM scenarios,
where the inclusion of additional degrees of freedom leads to an
enlargement of the flavor symmetry and thus to additional $U(1)$
factors whose SSB result in remnant discrete symmetries. This includes
models with larger symmetry groups and models with a
non-trivial embedding of the quark sector. \\

{\bf Model I:} With a setup defined as above, one can envisage
different flavor-invariant Lagrangians depending on the number of
available gauge singlet flavon fields.  Assuming a minimal content,
subject to flavor transformation properties given by $Y_e=(3,\bar
3,1)_{(1,-1,0,-1)}$, $Y_\nu=(3,1,\bar 3)_{(1,0,0,1)}$ and
$\sigma=(1,1,\bar 6)_{(0,0,-2,0)}$, the following lepton sector Yukawa
Lagrangian can be written:
\begin{equation}
  \label{lagrangian-MFV-nu-scot}
  {\cal L} =
  \frac{\lambda_e}{\Lambda}\overline{ \ell_{L}}\,Y_e\,e_{R}\,H
  +
  \frac{\lambda_\nu}{\Lambda}\overline{ \ell_{L}}\,Y_\nu\,\nu_{R}\, \Phi
  +
  \lambda_\sigma \, \overline{\nu^c_R} \, \sigma\, \nu_R + \mbox{H.c.}\ .
\end{equation}
Here, one can also choose $\sigma$ to be an $SU(3)_{\nu_R}$
triplet. Such a choice will certainly affect the Majorana mass
spectrum, but will not have any impact in our conclusions. Note that
one cannot write charged lepton Yukawa couplings involving $\Phi$, nor
type-I seesaw Yukawa couplings involving $H$. Their presence would
require extra ``Yukawa'' fields, which are absent as demanded by our
minimality criteria. The flavor-invariant and renormalizable scalar
potential consist of three pieces:
\begin{equation}
  \label{eq:full-scalar-potential-Z2-case}
  V= V_\text{SM} + V(H,\Phi) + V(H,\Phi,\sigma)\ ,
\end{equation}
where $V_\text{SM}$ has an obvious meaning and $V(H,\Phi,\sigma)$
involves quadratic and quartic $\sigma$ terms as well as mixed
$H-\sigma$ and $\Phi-\sigma$ terms. Of particular interest for
neutrino mass generation is the $V(H,\Phi)$ piece, which we explicitly
write:
\begin{equation}
  \label{scalar-potential-H-Phi-1}
  V(H,\Phi)= M^2_\Phi \left|\Phi\right|^2 
  + 
  \lambda_\Phi\left|\Phi\right|^4
  +
  \lambda_{H\Phi}\left|H \right|^2\left|\Phi\right|^2
  +
  \lambda_a \left|H^T \epsilon\, \Phi\right|^2 \ .
\end{equation}
As long as $V(H,\Phi,\sigma)$ allows for a $U(1)_{\nu_R}\times
SU(3)_{\nu_R}$ non-invariant ground state, the $U(1)_{\nu_R}\times
SU(3)_{\nu_R}$ symmetry will be spontaneously broken to $\mathbb{Z}_2$
via $\langle\sigma\rangle \ne 0$ \footnote{Depending on the vev
  structure of $\langle\sigma\rangle$ one could also have an
  additional remnant symmetry e.g. if $\langle\sigma\rangle \propto
  I_{3\times 3}$, one will be left with a residual $O(3)_{\nu_R}$.}.
The $\mathbb{Z}_2$ Abelian discrete symmetry is, therefore, a residual
symmetry resulting from the SSB of the global $U(1)_{\nu_R}$ factor.
At this symmetry breaking stage, $\mathbb{Z}_2$ is an exact symmetry
of the full Lagrangian under which the SM fields are even while the
BSM fields $\nu_R$ and $\Phi$ are odd. It will remain so---even after
electroweak symmetry breaking---provided $M^2_\Phi$ and $\lambda_{a}$
in (\ref{scalar-potential-H-Phi-1}) are positive, case in which
$\langle\Phi\rangle=0$.

\begin{figure}[t]
  \centering
  \includegraphics[scale=0.8]{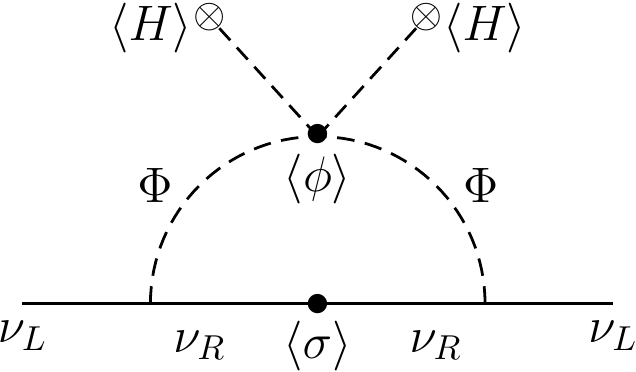}
  \caption{Feynman diagram responsible for the neutrino mass matrix in
    the $\mathbb{Z}_2$-based model.}
  \label{fig:1loop-diagram}
\end{figure}                    %

Under these conditions, the setup of
eqs. (\ref{lagrangian-MFV-nu-scot}),
(\ref{eq:full-scalar-potential-Z2-case}) and
(\ref{scalar-potential-H-Phi-1}) is $U(1)_L$ invariant (even though
right-handed neutrino Majorana mass terms are generated after SSB of
$U(1)_{\nu_R}\times SU(3)_{\nu_R}$). Since light Majorana neutrino
masses demand lepton number violation, new terms are then
required. The simplest choice is to include a new flavon
$\phi=(1,1,1)_{(0,0,-2,0)}$ which enables extending the scalar
potential in (\ref{eq:full-scalar-potential-Z2-case}) with a
non-renormalizable term:
\begin{equation}
  \label{eq:non-renormalizable-term-scotogenic}
  V(H,\Phi)\supset
  \frac{\lambda_\phi}{\Lambda} \, 
  \phi\left(H^T \epsilon\, \Phi\right)^2 + \mbox{H.c.}\ .
\end{equation}
After $\phi$ acquires a vev, this term will induce an effective
coupling $\lambda_5\left(H^T \epsilon\, \Phi\right)^2$, with
$\lambda_5=\lambda_\phi\langle\phi\rangle/\Lambda$. The presence of
this new term implies unavoidably $U(1)_L$ breaking (by two
units). The setup therefore generates light Majorana neutrino masses
at the 1-loop level via the exchange of right-handed neutrinos and the
neutral CP-even and CP-odd components of $\Phi$
($\mathbb{R}\mbox{e}(\Phi^0)$ and $\mathbb{I}\mbox{m}(\Phi^0)$,
respectively), as shown in Fig. \ref{fig:1loop-diagram} and exactly as
in the scotogenic model \cite{Ma:2006km}. Note that $\lambda_5$ being
an effective coupling, it is expected to be small
($\langle\phi\rangle\ll \Lambda$), which in turn implies a small mass
splitting between $\mathbb{R}\mbox{e}(\Phi^0)$ and
$\mathbb{I}\mbox{m}(\Phi^0)$. \\

{\bf Model II:} We now go a step further and consider the case of a
remnant higher order cyclic symmetry, where not only the would-be
Yukawa couplings are effective terms, but also the
$\overline{\nu^c_R}\nu_R$ coupling.  This requires the introduction of
three flavon fields, $\rho=(1,1,\bar 6)_{(0,0,-3,0)}$,
$\phi_1=(1,1,1)_{(0,0,1,0)}$ and $\phi_2=(1,1,1)_{(0,0,-3,0)}$. In
addition, we also include three generations of the new fermionic field
$N_L$, the left-handed Dirac partner of $\nu_R$, with exactly the same
transformation properties under all gauge and flavor
symmetries~\footnote{As we will see in the following, without $N_L$,
  $\nu_R$ would be a massless dark matter particle, something ruled
  out by cosmological data. Moreover, this is actually required for
  gauge anomaly cancellation if we promote e.g. $U(3)_{\nu_R}$ to a
  local symmetry.}. Under these considerations, the Yukawa sector
consists of the first two non-renormalizable terms
in~(\ref{lagrangian-MFV-nu-scot}) and the terms
\begin{equation}
  \label{eq:non-renormalizable-Maj-mass-term}
  {\cal L}\supset
  \frac{\lambda_{\phi\rho}}{\Lambda} \, \phi_1\overline{N^c}\,\rho\,N 
  + M_N \overline{N} N
  + \mbox{H.c.} \,,
\end{equation}
where $N = N_L + \nu_R$. Flavor invariance allows, of course, for
scalar terms which are readily derivable, and so we do not write them
explicitly. We just highlight the existence of two terms of particular
relevance for neutrino mass generation:
\begin{equation}
  \label{eq:lagrangian-MFV-scalar-Z3}
  V\supset
  \lambda_{12}\, \phi_1^3 \phi_2
  +
  \mu_{\Phi} \, H^T \epsilon\, \Phi \,\phi_1 
  + 
  \mbox{H.c.}\ .
\end{equation}
We will assume that the complete scalar potential has a minimum
characterized by $\langle\Phi\rangle = \langle\phi_1\rangle=0$,
$\langle \rho \rangle \neq 0$ and $\langle \phi_2 \rangle \neq
0$. Thus, in this case, $U(1)_{\nu_R} \times SU(3)_{\nu_R}$ gets
spontaneously broken, leaving a remnant $\mathbb{Z}_3$ symmetry under
which $N \to e^{2\pi i/3}\,N$, $\Phi \to e^{4\pi i/3}\,\Phi$,
$\phi_1 \to e^{2\pi i/3}\,\phi_1$ and the remaining fields transform
trivially. As in the previous example, this symmetry is generated from
$U(1)_{\nu_R}$ SSB.

\begin{figure}[t]
  \centering
  \includegraphics[scale=0.8]{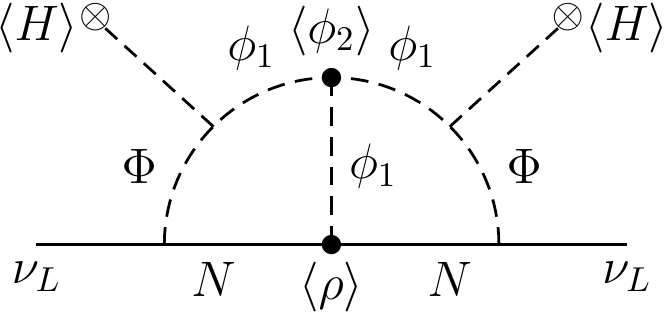}
  \caption{Feynman diagram responsible for the neutrino mass matrix in
    the $\mathbb{Z}_3$-based model.}
  \label{fig:2loop-diagram}
\end{figure}                    %

In the present setup $U(1)_L$ is violated, inducing Majorana neutrino
masses at the 2-loop order, as depicted in
Fig. \ref{fig:2loop-diagram}.  In the same vein of the $\mathbb{Z}_2$
case, the remnant $\mathbb{Z}_3$ symmetry allows for dark matter
stabilization, which can be either of fermionic or scalar nature,
namely $N$, $\Phi^0$ or $\phi_1$. However, in contrast to the usual
$\mathbb{Z}_2$-based dark matter scenarios, this dark matter particle
will have semi-annihilation
processes~\cite{Hambye:2008bq,D'Eramo:2010ep}.

\section{Conclusions}
\label{sec:conclusions}%
We have pointed out that the same dynamical flavor symmetry that
governs SM fermions mass hierarchies and mixing patterns, might be as
well at the origin of Abelian discrete symmetries,
$\mathbb{Z}_N$. Spontaneous symmetry breaking of the Abelian sector
yields such symmetries, provided the flavored UV completion involves
flavon fields with suitable charges. These symmetries, which quite
often are {\it ad hoc} ``artifacts'', are employed for Majorana
neutrino mass generation and dark matter stabilization, among
others. Thus, we have suggested that the discrete symmetry generated
in this way, provides a non-trivial link between the {\it theory} of
flavor and the origin of neutrino masses and dark matter. We have
shown the feasibility of this approach by constructing 
$\mathbb{Z}_2$- and $\mathbb{Z}_3$-based models, the former resembling
the well-known scotogenic model, while the latter a new realization
with quite a few interesting phenomenological implications.

In summary, we argued that discrete $\mathbb{Z}_N$ symmetries have a
dynamical flavor origin, and we have illustrated how this approach can
be implemented. Finally, we would like to stress that this picture
offers several interesting theoretical as well as phenomenological
avenues which are worth exploring.


\section*{Acknowledgements}
CFS is supported by Funda\c{c}\~ao de Amparo \`a Pesquisa do Estado de
S\~ao Paulo (FAPESP). He acknowledges the hospitality of
``Universit\'e de Li\`ege'' when the idea of this paper was being
conceived.  DAS is supported by a ``Charg\`e de Recherches'' contract
funded by the Belgian FNRS agency and would like to thank Enrico Nardi
for enlightening discussions about the SM flavor problem. AV
acknowledges partial support from the EXPL/FIS-NUC/0460/2013 project
financed by the Portuguese FCT.

\end{document}